# Concept and optical design of the cross-disperser module for CRIRES+

E. Oliva*[a], A. Tozzi[a], D. Ferruzzi[a], L. Origlia[b], A. Hatzes[c], R. Follert[c], T. Loewinger[c], N. Piskunov[d], U. Heiter[d], M. Lockhart[d], T. Marquart[d], E. Stempels[d], A. Reiners[e], G. Anglada-Escude[e], U. Seemann[e], R. J. Dorn[f], P. Bristow[f], D. Baade[f], B. Delabre[f], D. Gojak[f], J. Grunhut[f], B. Klein[f], M. Hilker[f], D. J. Ives[f], Y. Jung[f], H.U. Kaeufl[f], F. Kerber[f], J.L. Lizon[f], L. Pasquini[f], J. Paufique[f], E. Pozna[f], A. Smette[f], J. Smoker[f], E. Valenti[f]

[a] INAF-Arcetri Observatory, largo E. Fermi 5, I-50125 Firenze, Italy;
[b] INAF-Bologna Observatory, via Ranzani 1, I-40127 Bologna, Italy;
[c] Thueringer Landessternwarte, Sternwarte 5, D-07778 Tautenburg, Germany;
[d] Uppsala University Physics and Astronomy, Box 515, 751 20 Uppsala, Sweden;
[e] Goettingen University, Institute for Astrophysics, F. Hund Platz 1, D-37077 Goettingen, Germany;
[f] ESO, K. Schwarzschild str. 2; D-85748 Garching bei Muenchen, Germany

## ABSTRACT

CRIRES, the ESO high resolution infrared spectrometer, is a unique instrument which allows astronomers to access a parameter space which up to now was largely uncharted. In its current setup, it consists of a single-order spectrograph providing long-slit, single-order spectroscopy with resolving power up to R=100,000 over a quite narrow spectral range. This has resulted in sub-optimal efficiency and use of telescope time for all the scientific programs requiring broad spectral coverage of compact objects (e.g. chemical abundances of stars and intergalactic medium, search and characterization of extra-solar planets). To overcome these limitations, a consortium was set-up for upgrading CRIRES to a cross-dispersed spectrometer, called CRIRES+. This paper presents the updated optical design of the cross-dispersion module for CRIRES+. This new module can be mounted in place of the current pre-disperser unit. The new system yields a factor of >10 increase in simultaneous spectral coverage and maintains a quite long slit (10"), ideal for observations of extended sources and for precise sky-background subtraction.

## 1. INTRODUCTION

The cryogenic high resolution IR echelle spectrograph (CRIRES[1]) is the ESO infrared (0.95-5.4 microns) high resolution spectrograph operating at the Nasmyth A focus of VLT-UT1. The instrument provides long-slit (40") spectroscopy of a single order of the echelle grating. The maximum resolving power is R=100,000 and the spectral coverage is quite narrow, about 1/70 to 1/50 of the central wavelength. Observations of compact objects (e.g. stellar photospheres) could be made much more efficient by implementing a cross-dispersed mode, which increases the simultaneous spectral coverage by one order of magnitude or more. To overcome these limitations, we set up a consortium for upgrading CRIRES to a cross-dispersed spectrometer, called CRIRES+. In the early phase of the project we assumed that CRIRES+ would retain the same detectors, i.e. a linear mosaic of four InSb arrays[3] with 4096 x 512 usable pixels (effective area 111 x 14 mm). Consequently, the first design[2] of the cross-disperser unit did not require fundamental changes of the pre-disperser optics and was developed trying to minimize the changes. In practice, it consisted of adding a cryogenic grism wheel to the current pre-disperser unit of CRIRES. The strategy drastically changed after we secured the funds necessary to upgrade the detector with a mosaic of three HAWAII-2 arrays that increase the size of the focal plane to 6144 x 2048 pixels (effective area 111 x 37 mm, a factor 2.7 larger in the direction perpendicular to the main dispersion).

* oliva@arcetri.astro.it; phone +39 0552752291

Luckily, the main spectrometer unit of CRIRES is sufficiently oversized to feed the new focal plane without any significant deterioration of image quality or vignetting (see Sect. 4). Instead, the cryogenic pre-disperser unit currently mounted in CRIRES cannot handle such an increase of slit-length. We therefore designed a new cross-disperser unit that fits within the same volume, with minimum impacts on the cryo-mechanics, electronics and controls of the instrument. This paper provides a detailed description of the optical design of this unit. Other aspects of CRIRES+ are addressed in a series of papers[4,5,6,7,8]

## 2. OVERALL OPTICAL CONCEPT AND DESIGN

Figure 1 summarizes the overall concept of the CRIRES to CRIRES+ upgrade. The main, high resolution spectrometer unit remains untouched. The new cross-disperser unit substitutes the current re-imager and pre-dispersing sub-systems.

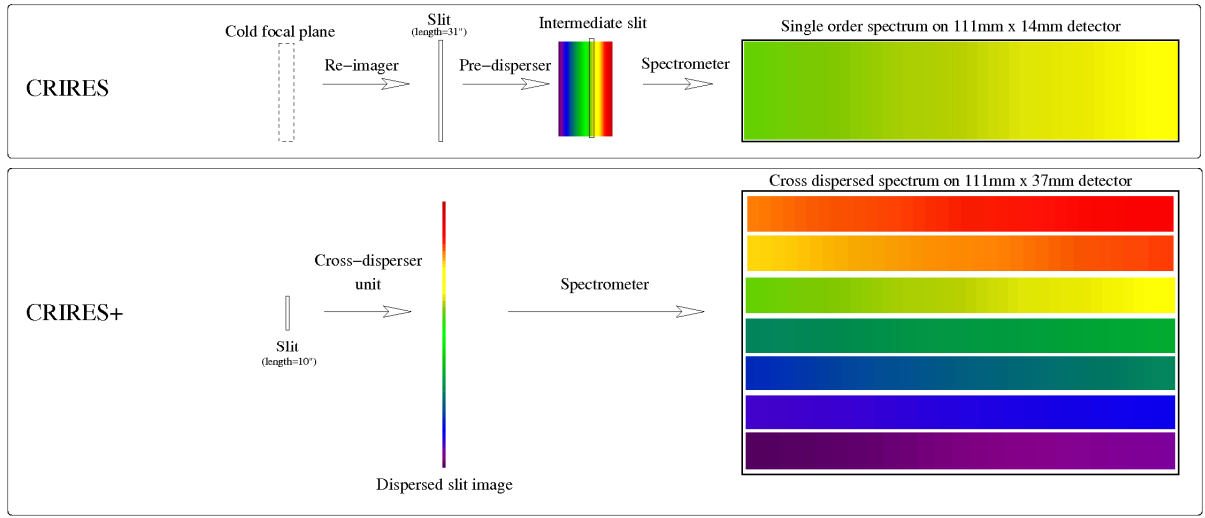

**Figure 1** Schematic diagram summarizing the differences between CRIRES and CRIRES+.

The optical layout is shown in Figure 2. The design was performed starting from the current (as-built) elements of the cold (spectrometer + echelle grating) and warm AO adapter[9] optics. Following the optical path (left to right) one finds the following elements.
- The cryostat window, also acting as the dichroic beam splitter that feeds the visible light to the adaptive optics wave-front sensor (not shown in Figure 2). The new window is similar to that currently mounted but with a new coating to improve the transmission in the bluer part of the Y band (0.97-1.00 microns).
- Two interchangeable slits (width=0.2" and 0.4", length=10") at the F/15 focus. Each slit is cut into a mirror reflecting the outer light to the guider optics.
- The guider optics, marked with "G" in Figure 1. They have a field of view of 20"x40" (sky-projected angles) and cover the 1-2.5 microns wavelength range. They consist of
    - A mildly spherical mirror, acting as pupil re-imager (GM1).
    - A flat mirror, acting as folder (GM2).
    - A lens-doublet camera, which creates an image of the entrance field.
    - A few interchangeable filters. We may re-use those already mounted in the CRIRES guider.
    - One IR-detector. We may re-use the one already mounted in the CRIRES guider.
- An off-axis parabolic mirror, which collimates the light passing through the slit. The diameter of the collimated beam is 50mm.
- A flat mirror (Flat1) folding the beam.
- Three selectable long-pass filters mounted on a wheel. They are needed to cut the 2$^{nd}$ and higher orders of the cross-disperser gratings.
- Another flat mirror (Flat2) that folds the beam.

- Six selectable gratings that disperse the light in the direction perpendicular to the dispersion of the main spectrometer. These are the cross-dispersers.
- A lens-triplet camera, achromatic over the full 0.95-5.4 microns range. It creates the dispersed image of the slit at the entrance of the spectrometer. The beam aperture at the focus is F/8. To minimize stray-light, a field stop (i.e. a slightly oversized slit) can be inserted at this position.
- A flat mirror (Spec-Flat1), which substitutes the current pupil-reimaging mirror.
- The spectrometer optics, consisting of 3 mirrors (TMA1, TMA2, TMA3) used in double pass and one echelle grating mounted on a rotating stage. These elements are the same as in the original CRIRES.

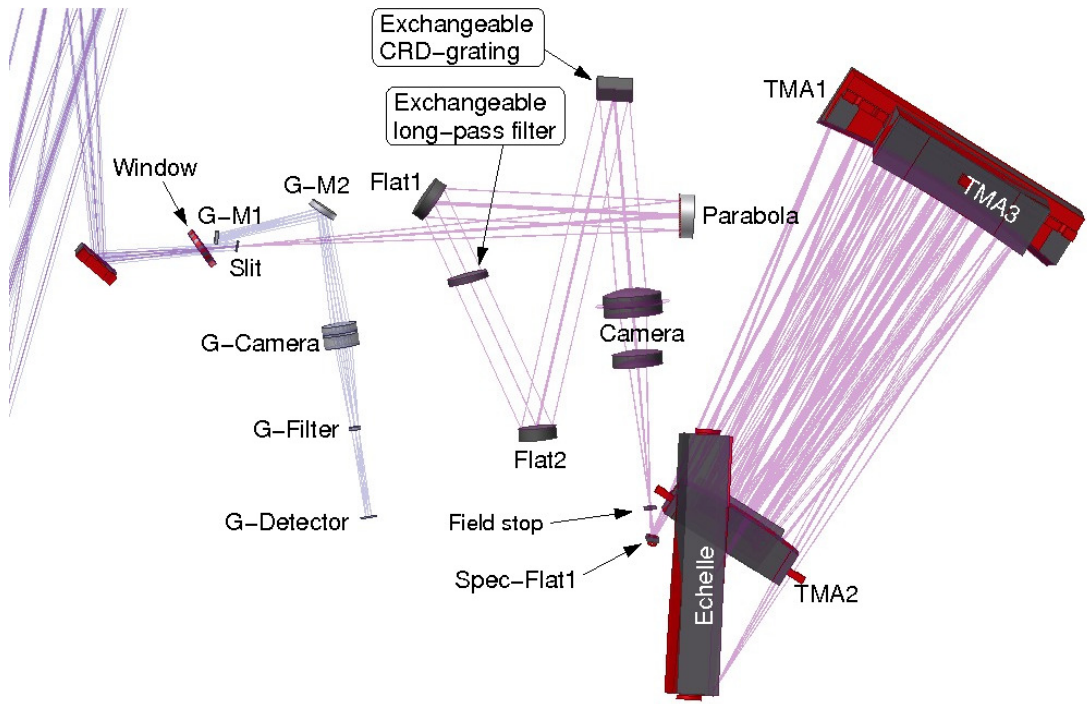

**Figure 2** Optical layout and rays tracing for CRIRES+

## 3. SPECTRAL FORMAT AND WAVELENGTH COVERAGE

The parameters of the cross-dispersing gratings are designed to cover one photometric band per grating, avoiding the regions of bad telluric transmission. Therefore, a spectrum taken with a given cross-disperser-grating includes the orders falling within the corresponding photometric band. The observing bands and corresponding orders are summarized in Table 1. This table also includes the parameters of ideal cross-dispersing gratings and of order-sorter filters. Several of the gratings parameters are quite close to those of commercial devices. These off-the-shelf devices will be acquired, tested and characterized in a configuration similar to CRIRES+. Those gratings passing the CRIRES+ technical specifications will then be used in the instruments. Custom gratings will be purchased for the bands for which no suitable commercial gratings are available.

The parameters of the order-sorter filters have been selected to minimize their number. This results in only 3 filters. Each of them can be used in combination with two gratings.

The layouts of the spectra are displayed in Figure 3. The echellograms are computed with the echelle angle at its blaze position. The wavelengths falling outside of the detector in the left/right direction can be accessed by changing the angle of the echelle grating. The number of settings necessary to get full spectral coverage is given in the fourth column of Table 1.

**Table 1** Main parameters of CRIRES+ cross-dispersed spectra. Slit length is always 10" (sky-projected angles).

| Band | λ–range (µm) | Orders | $N_{exp}$[1] | Grating Angle | Grating Lines/mm | Filter |
|---|---|---|---|---|---|---|
| Y | 0.96 – 1.13 | 50 – 59 | 1 | 14.8° | 480 | Cut-on 800nm - RG850[2] |
| J | 1.13 – 1.36 | 42 – 50 | 2 | 13.2° | 360 | Cut-on 800nm - RG850[2] |
| H | 1.48 – 1.81 | 31 – 38 | 2 | 11.0° | 230 | Cut-on 1350nm |
| K | 1.93 – 2.54 | 22 – 29 | 3 | 8.5° | 130 | Cut-on 1350nm |
| L | 2.85 – 4.20 | 14 – 20 | 4 | 6.2° | 60 | Cut-on 2800nm |
| M | 3.58 – 5.50 | 10 – 16 | 5 | 5.4° | 40 | Cut-on 2800nm |

*Notes to table*
[1] Number of exposures needed to cover the full band. These spectra must be taken with the echelle grating positioned at different angles. Extra positions are needed to fill the gaps between the focal-plane-mosaic of detector arrays.
[2] RG850 is a standard, colored glass. The filter could just be a disk of this glass, A/R coated.

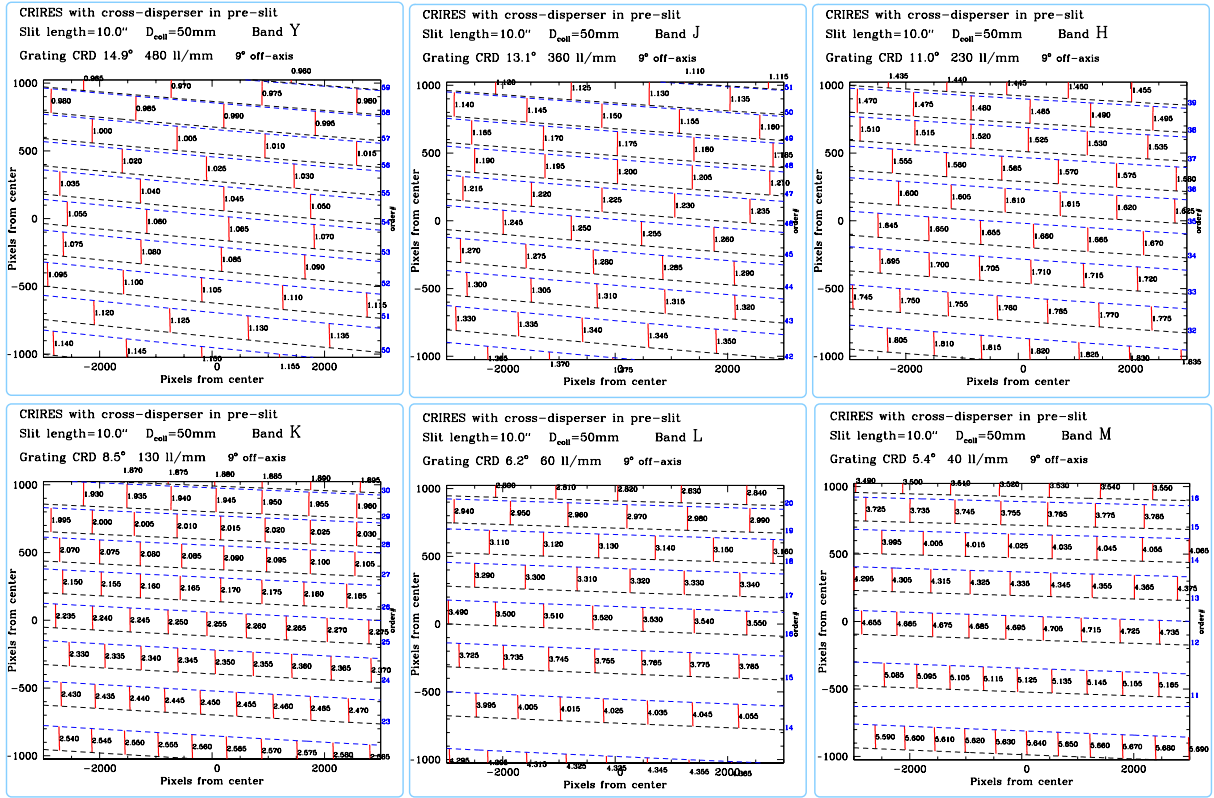

**Figure 3** Layout of the CRIRES+ spectra with the echelle grating at its blaze position.

## 4. OPTICAL PERFORMANCES

The spots diagrams of the complete CRIRES+ system (from the telescope focus to detector) are shown in Figure 4. The size of the squares is 60 microns. This corresponds to the width of the narrowest slit (0.2"). The circles represent the FWHM of the diffraction peak. The spots are shown for the lowermost, central and uppermost orders. For each order, the spots are shown for three positions along the 10" slit, and for three wavelengths. The bottom, central and upper rows refer to the lower edge, center and upper edge of the slit, respectively. The left, central and right-hand columns refer to the bluest, central, and reddest wavelengths.

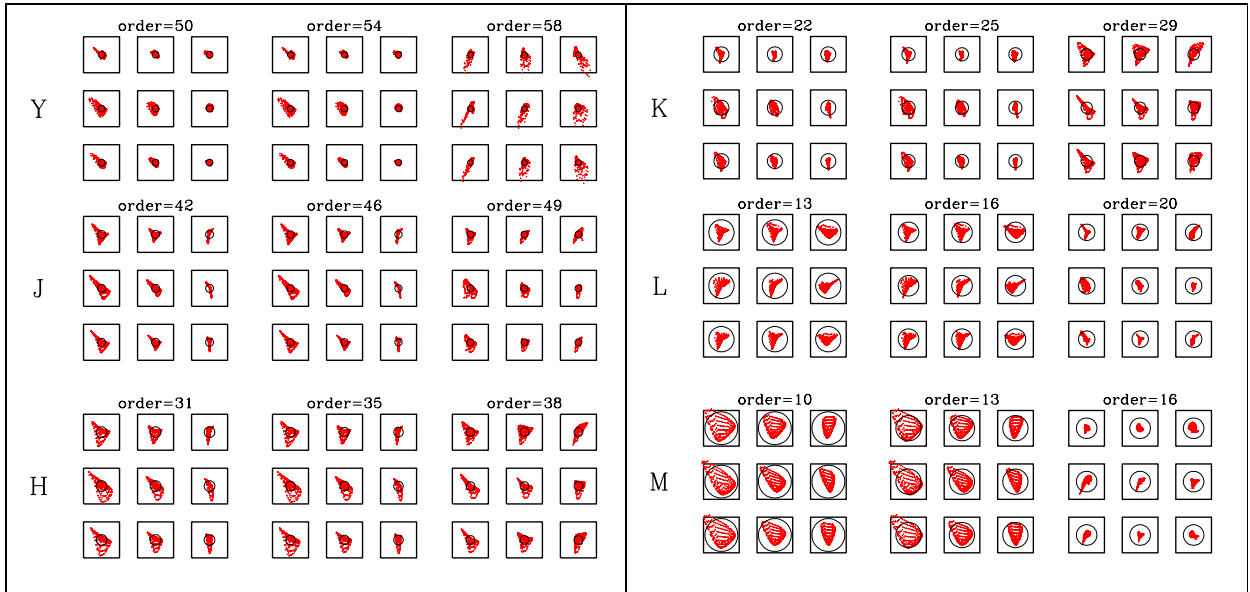

. **Figure 4** Spots diagrams for CRIRES+, see the first paragraph of Sect. 4 for details.

The illumination of the (already existing) spectrometer mirrors is visualized in Figure 5. The footprints of the rays are always within the clear apertures of the mirrors.

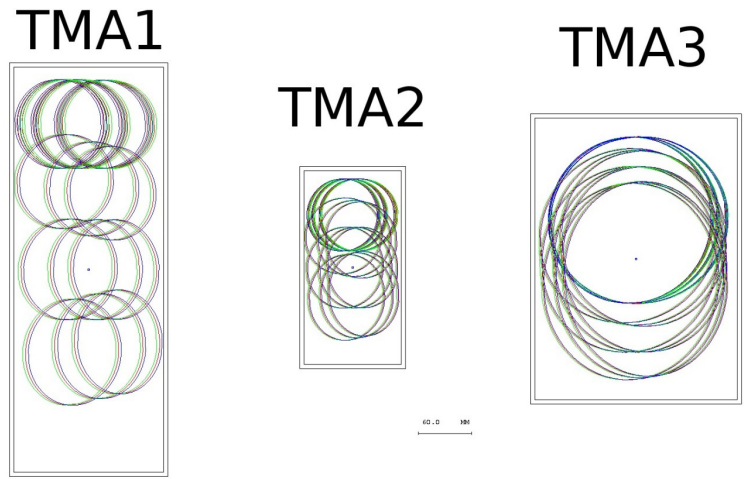

**Figure 5** Footprints of the rays, in the cross-dispersed mode, on the already existing TMA mirrors of the CRIRES spectrograph.

# 5. DETAILS OF THE NEW OPTICAL ELEMENTS

Figure 6 shows the apertures and illumination footprints of the optical elements of the cross-disperser module. Their parameters are described in Sections 5.1 to 5.3. The optical elements for the guider are described in Section 5.4.

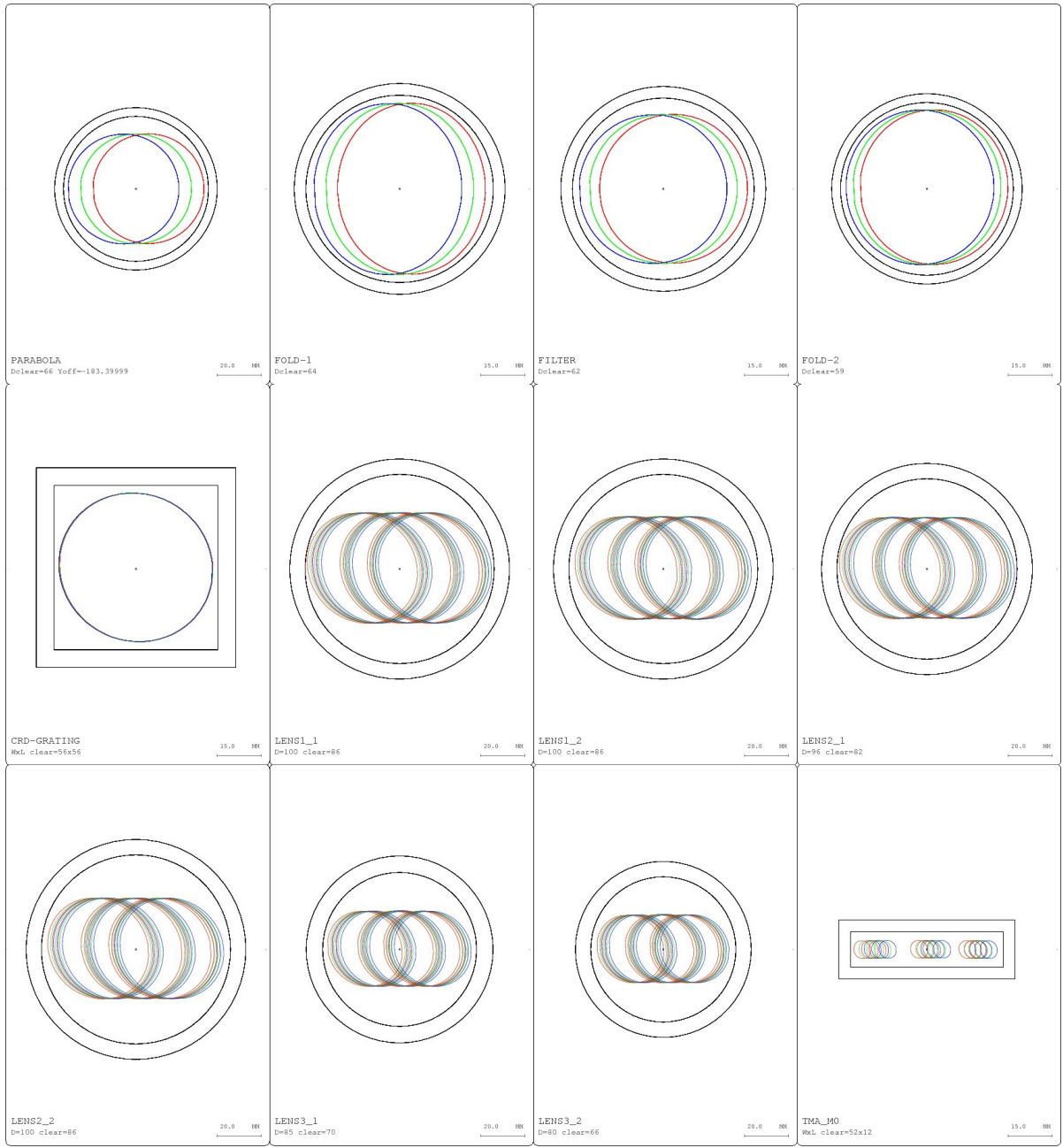

**Figure 6** Illumination footprints and clear apertures for the new optical elements of CRIRES+

## 5.1 Mirrors

The main parameters of the mirrors are summarized in Tables 2 to 8. All dimensions are in mm and at room temperature. Positioning tolerances are estimated assuming that the focusing of the lens-triplet camera is the only compensator.

**Table 2** Common parameters for all mirrors

| Parameter | Value | Comment |
| --- | --- | --- |
| Mirror substrate | Aluminum 6061, cryogenically aged | Al-6082 could also be used |
| Surface figure | <20 nm, rms | Over the full clear aperture |
| Surface roughness | <5 nm, rms | |
| Coating | Gold | |
| Positioning tolerances | ± 0.1 mm   ± 3 arcmin | |

**Table 3** Collimator mirror

| Parameter | Value | Comment |
| --- | --- | --- |
| Curvature radius | 1500 ± 0.5% (goal ± 0.3%) | |
| Conic constant | -1.0 | Parabola |
| Clear aperture | D=66  de-centered by 91.7 ± 0.1 | De-center relative to vertex |
| Outer diameter | 74 | Indicative value, TBC |
| Thickness at vertex | 20 | Indicative value, TBC |

**Table 4** Flat-1 mirror

| Parameter | Value | Comment |
| --- | --- | --- |
| Curvature radius | Flat | |
| Clear aperture | D=64 | |
| Outer diameter | 72 | Indicative value, TBC |
| Thickness | 20 | Indicative value, TBC |

**Table 5** Flat-2 mirror

| Parameter | Value | Comment |
| --- | --- | --- |
| Curvature radius | Flat | |
| Clear aperture | D=59 | |
| Outer diameter | 65 | Indicative value, TBC |
| Thickness | 20 | Indicative value, TBC |

**Table 6** Spec-flat 1 mirror (at the entrance of TMA spectrometer, TMA-M0 in Fig. 6)

| Parameter | Value | Comment |
| --- | --- | --- |
| Curvature radius | Flat | |
| Clear aperture | 52 x 12 | Rectangular aperture |
| Outer size | 60 x 20 | Indicative value, TBC |
| Thickness | 10 | Indicative value, TBC |

**Table 7** Field mirror for guider optics (GM1 in Fig. 2)

| Parameter | Value | Comment |
| --- | --- | --- |
| Curvature radius | 1320 ± 0.5% | |
| Clear aperture | 32 x 16 | Rectangular aperture |
| Outer size | 40 x 20 | Indicative value, TBC |
| Thickness | 5 | Indicative value, TBC |

**Table 8** Fold mirror for guider optics (GM2 in Fig. 2)

| Parameter | Value | Comment |
|---|---|---|
| Curvature radius | Flat | |
| Clear aperture | D=50 | |
| Outer diameter | 60 | Indicative value, TBC |
| Thickness | 10 | Indicative value, TBC |

**5.2 Dichroic window and order-sorter filters**

A new dichroic window is necessary because of the TLR requirement on spectral coverage in the Y-band (0.96-1.13 microns). The current window has a very low transmission below 1 micron and, therefore, needs to be replaced. The order-sorter filters are needed to block the short-wavelength light which, diffracted to 2$^{nd}$ and higher orders of the CRD grating, may contaminate the spectrum. The main parameters are summarized in Tables 9 to 13. All dimensions are in mm and at room temperature.

**Table 9** Dichroic window

| Parameter | Value | Comment |
|---|---|---|
| Substrate | CaF$_2$ | |
| Clear aperture | 63 | |
| Outer diameter | 72 +0/-0.2 | |
| Central thickness | 10 ± 0.1 | |
| Wedge | 1 deg ± 6 arc-min | |
| Surface figure | 30 nm rms | Over clear aperture, focus removed |
| Surface roughness | 5 nm rms | |
| Incidence angle of light | Cone axis tilted by 30 deg<br>Total cone aperture 4 deg | For dichroic design |
| Coating on first surface | Dichroic,<br>R>90% over 450nm – 930nm<br>T>90% over 960nm – 1350nm<br>T>90% over 1470nm – 1830nm<br>T>90% over 1950nm – 2500nm<br>T>90% over 2900nm – 4200nm<br>T>90% over 4500nm – 5300nm | R, T intended as average values in given bands. |
| Coating on second surface | Anti-reflection<br>T>99% over the same transmission bands of the dichroic surface | |
| Lateral mark on window | Arrow or similar | To indicate which of the two surfaces is the dichroic |

**Table 10** Order sorter filter #1a (for Y, J bands), first option with colored glass

| Parameter | Value | Comment |
|---|---|---|
| Substrate | RG850 | |
| Clear aperture | 62 | |
| Outer diameter | 70 | Indicative value, TBC |
| Central thickness | 3 | Standard value for colored glasses |
| Wedge | <5 arc-sec | Deviation of ray <3 arc-sec |
| Surface figure | 50 nm rms | Over clear aperture, including focus |
| Surface roughness | 5 nm rms | |
| Incidence angle of light | Cone axis tilted by 10 deg<br>Total cone angle 1 deg | For coating design |
| Coating on both surface | Anti-reflection<br>R<0.5% over 960nm – 1350nm | R is average reflection per surface in given band. |

**Table 11** Order sorter filter #1b (for Y, J bands), second option with dedicated coating

| Parameter | Value | Comment |
|---|---|---|
| Substrate | Infrared grade fused silica | |
| Type of filter | Single element | A sandwich could be considered if this reduces costs |
| Clear aperture | 62 | |
| Outer diameter | 70 | Indicative value, TBC |
| Central thickness | 10 | Indicative value, TBC |
| Wedge | <5 arc-sec | Deviation of ray <3 arc-sec |
| Surface figure | 50 nm rms | Over clear aperture, including focus |
| Surface roughness | 5 nm rms | |
| Incidence angle of light | Cone axis tilted by 10 deg<br>Total cone angle 1 deg | For coating design |
| Filter transmission | T<0.01% over 350nm – 800nm<br>T>95% over 960nm – 1350nm | T is the average transmission of the whole filter in given bands. |

**Table 12** Order sorter filter #2 (for H,K bands)

| Parameter | Value | Comment |
|---|---|---|
| Substrate | Infrared grade fused silica | |
| Type of filter | Single element | A sandwich could be considered if this reduces costs |
| Clear aperture | 62 | |
| Outer diameter | 70 | Indicative value, TBC |
| Central thickness | 10 | Indicative value, TBC |
| Wedge | <5 arc-sec | Deviation of ray <3 arc-sec |
| Surface figure | 50 nm rms | Over clear aperture, including focus |
| Surface roughness | 5 nm rms | |
| Incidence angle of light | Cone axis tilted by 10 deg<br>Total cone angle 1 deg | For dichroic design |
| Filter transmission | T<0.01% over 350nm – 1270nm<br>T>95% over 1470nm – 1830nm<br>T>95% over 1950nm – 2500nm | T is the average transmission of the whole filter in given bands. |
| Coating design | Dichroic on $1^{st}$, A/R on $2^{nd}$ surface | Suggested, TBC |

**Table 13** Order sorter filter #3 (for L,M bands)

| Parameter | Value | Comment |
|---|---|---|
| Substrate | $CaF_2$ | Other materials transparent over 2800nm – 5400nm could be considered |
| Type of filter | Single element | A sandwich could be considered if this reduces costs |
| Clear aperture | 62 | |
| Outer diameter | 70 | Indicative value, TBC |
| Central thickness | 10 | Indicative value, TBC |
| Wedge | <5 arc-sec | Deviation of ray <3 arc-sec |
| Surface figure | 50 nm rms | Over clear aperture, including focus |
| Surface roughness | 5 nm rms | |
| Incidence angle of light | Cone axis tilted by 10 deg<br>Total cone angle 1 deg | For dichroic design |
| Filter transmission | T<0.01% over 350nm – 2700nm<br>T>90% over 2900nm – 4200nm<br>T>90% over 4500nm – 5300nm | T is the average transmission of the whole filter in given bands. |

### 5.3 Lens-triplet camera

The layout of the camera is shown in Figure 7 and the main parameters are summarized in Table 14. All linear dimensions are in mm. The central thickness (CT) and curvature radii (R) are at ambient temperature. They have been scaled from their optimized values at cryogenic temperatures using the following integrated coefficients of thermal expansion:

$R(BaF_2, 20\ ^oC) = 1.00322\ R(BaF_2, 70\ K)$
$R(ZnSe, 20\ ^oC) = 1.00119\ R(ZnSe, 70\ K)$
$R(LiF_2, 20\ ^oC) = 1.00494\ R(LiF_2, 70\ K)$

The conic constant (K) is temperature-independent. The central distances between the lenses are at 70 K.

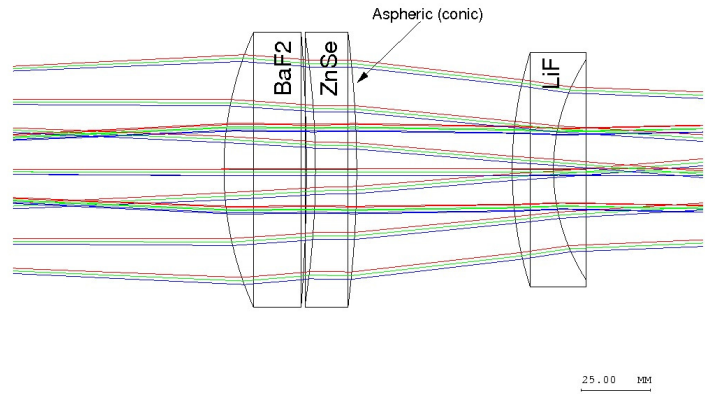

**Figure 7** Layout and rays-tracing of the lens-triplet camera

**Table 14** Parameters of the lens-triplet camera.

| Element | Material | CT | $D_{out}$ (indicative values) | $D_{clear}$ | R | K | comment |
|---|---|---|---|---|---|---|---|
| Lens 1 (biconvex) | $BaF_2$ | 30.1 | 100.0 | >86 | 121.36 | - | convex |
| | | | | >86 | 648.98 | - | convex |
| Lens 2 (meniscus) | ZnSe | 15.0 | 96.0 | >82 | 321.91 | - | concave |
| | | | 100.0 | >86 | 387.09 | -6.465 | convex |
| Lens 3 (meniscus) | $LiF_2$ | 15.1 | 85.0 | >70 | 142.12 | - | convex |
| | | | 80.0 | >66 | 73.71 | - | concave |
| Centeral distances between the optical elements (at 70 K) ||||||||
| Lens 1 – Lens 2 | | | 3.3 | Lens 2 – Lens 3 | | | 56.6 |
| Grating – Lens 1 | | | 330.0 | Lens 3 – Mirror TMA0 | | | 387.3 |

#### 5.3.1 Tolerances

Tolerances were computed with the conservative assumption of a "blind-mounting", i.e. an alignment procedure which does not foresee the possibility of separately adjusting the central distances, lateral positions (shifts along X,Y) and angles (tilts along α,β) of the lenses. We also assumed that the position along the optical axis (focus) of the whole camera can be actively modified at the operating temperature. The manufacturing tolerances are summarized in Table 15. The rms wave-front irregularity of each lens surface is measured in reflection over a diameter of 50mm

Table 15 Manufacturing tolerances for the lenses

| Element | Manufacturing tolerances | | | | | | |
|---|---|---|---|---|---|---|---|
| | Curvature | Conic coef. | C-Thickness | Wedge | Internal decenter | Surface figure (including power) | Surface roughness |
| Lens 1 (BaF$_2$) | 0.1% | - | 0.2 mm | 50 arc-sec | 0.02 mm | 60 nm rms | 5 nm rms |
| | 0.5% | - | | | | 60 nm rms | 5 nm rms |
| Lens 2 (ZnSe) | 0.1% | - | 0.2 mm | 20 arc-sec | 0.02 mm | 25 nm rms | 5 nm rms |
| | 0.3% | 0.5% | | | | 25 nm rms | 5 nm rms |
| Lens 3 (LiF$_2$) | 0.5% | - | 0.2 mm | 2 arc-min | 0.1 mm | 80 nm rms | 5 nm rms |
| | 0.2% | - | | | | 80 nm rms | 5 nm rms |

The manufacturing tolerances on curvatures can be relaxed to 0.5% (or even larger) by foreseeing the possibility of modifying the thickness of the spacer between the first (BaF$_2$) and the second (ZnSe) lenses. For example, a variation of 0.5% of the curvature of the most critical surfaces can be fully compensated by modifying the thickness of the L1-L2 spacer by about 1.5mm. A practical alignment procedure could be as follows
- Measure to an accuracy of <0.1% the actual curvature of the manufactured lenses.
- Compute with rays-tracings the optimal thickness of the L1-L2 spacer
- Build/assemble the camera holder using this spacer

The manufacturing tolerances on internal tilt and decenters can be relaxed to 5arc-min and 1mm by adding a mechanical adjustment of the centering and tilt of the first lens.

The camera assembly includes the lenses and is mounted on a sliding stage inside the spectrograph. The tolerances reported here refer to the positioning of the whole assembly relative to the rest of the optics (cross-disperser grating and spectrometer).

Table 16 Positioning tolerances of the whole camera, relative to the rest of the spectrograph

| Element | Lateral shift | Shift along optical axis | Tilt relative to optical axis |
|---|---|---|---|
| Camera assembly | 1 mm | Mechanically adjustable | 10 arc-min |

Table 17 lists the stability tolerances of the lenses for the two cameras. These tolerances also represent the accuracy of a mechanical holder where the lenses are blindly positioned, i.e. a situation where the alignment does not foresee the possibility of separately adjusting the shifts, tilts and central distances of the lenses.

Table 17 Positioning tolerances of the lenses within the camera

| Element | Stability and alignment tolerances | | |
|---|---|---|---|
| | Lateral shift | Shift along optical axis | Tilt |
| Lens 1 (BaF$_2$) | 0.03 mm | 0.1 mm | 2 arc-min |
| Lens 2 (ZnSe) | 0.03 mm | 0.1 mm | 2 arc-min |
| Lens 1 (LiF$_2$) | 0.05 mm | 0.2 mm | 3 arc-min |

### 5.3.2  Anti-reflection coating

All the surfaces must be coated to minimize reflection. The specification for the anti-reflection coating is R<1% (goal R<0.5%) per surface over the 960nm – 5400nm wavelengths range. Lower performances (i.e. higher reflection) may be accepted over the following ranges of wavelength

1350nm – 1470nm
1830nm – 1950nm
2500nm – 2900nm
4200nm – 4500nm

## 5.4 Positioning requirements for the cross-disperser gratings

The requirements on positioning repeatability of the cross-dispersing grating can be directly derived by the sensitivity coefficients listed in second row of Table 18. However, to avoid the risks of over-specifications, it is convenient to compare the effects of small angular changes of the cross-dispersing grating and of the echelle grating. For this reason we report, in the first row of Table 18, the sensitivity coefficients for the movements of the echelle grating. The most annoying effect on the data is the shift of the image along the spectrum, i.e. along the direction of dispersion of the echelle grating ($\Delta X$). This shift can be produced by a change of the echelle angle, or by a change of the angle of the cross-disperser in the direction perpendicular to its dispersion direction ($\Delta \alpha$). The sensitivity factors for these two effects are 14µm/arc-sec and 3.5µm/arc-sec, respectively. Therefore, a tilt of the cross-disperser grating produces an image-shift 4 times smaller than that produced by the same tilt of the echelle grating. Moreover, the tilt $\Delta \alpha$ is perpendicular to the direction of rotation of the cross-dispersers wheel (see Figure 8).

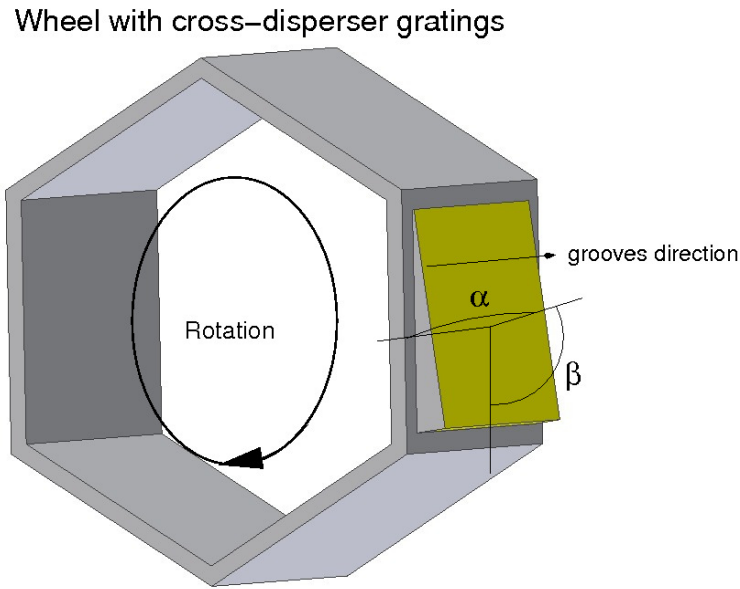

**Figure 8** Sketch of the wheel carrying the cross-disperser gratings

**Table 18** Sensitivity coefficients for tilts of the dispersing optical elements.

| Element | $\Delta Y/\Delta \alpha$ | $\Delta Y/\Delta \beta$ | $\Delta Y/\Delta \gamma$ | $\Delta X/\Delta \alpha$ | $\Delta X/\Delta \beta$ | $\Delta X/\Delta \gamma$ |
|---|---|---|---|---|---|---|
| Echelle grating | 0.03 | 6.86 | 13.1 | 13.8 | 1.14 | 0.10 |
| Cross-disperser grating | 0.15 | 3.92 | 0.04 | 3.52 | 0.05 | 0.92 |

*Notes to table*

All the sensitivity coefficients are in units of µm/arc-sec

$\Delta Y$ is the image shift parallel to cross-dispersion (perpendicular to echelle-dispersion)

$\Delta X$ is the image shift parallel to echelle-dispersion (perpendicular to cross-dispersion)

$\Delta \alpha$ is a tilt parallel to echelle-dispersion (perpendicular to cross-dispersion)

$\Delta \beta$ is a tilt parallel to cross-dispersion (perpendicular to echelle-dispersion)

$\Delta \gamma$ is a rotation of the grating around its nominal axis

## 5.5 Guider camera

The guider camera re-images the F/15 input focal plane onto the guider detector, at F/12. It has a field of view of 20"x40" (sky-projected angles) and is achromatic over the 1-2.5 microns wavelengths range. The layout of the guider camera is shown in Figure 9 and the main parameters are summarized in Table 19. All linear dimensions are in mm. The central thickness (CT) and curvature radii (R), are at ambient temperature. They have been scaled from their optimized values at cryogenic temperatures using the following integrated coefficients of thermal expansion:

$R(CaF_2, 20\ ^oC) = 1.00301\ R(CaF_2, 70\ K)$
$R(S\text{-}FTM16, 20\ ^oC) = 1.00167\ R(S\text{-}FTM16, 70\ K)$

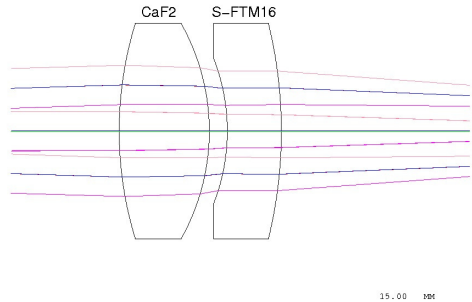

**Figure 9** Layout and rays-tracing of the guider camera

**Table 19** Parameters of the guider camera.

| Element | Material | CT | $D_{out}$ (indicative values) | $D_{clear}$ | R | comment |
|---|---|---|---|---|---|---|
| Lens 1 (biconvex) | $CaF_2$ | 25.0 | 70.0 | >50 | 102.52 | convex |
| | | | | >50 | 61.18 | convex |
| Lens 2 (meniscus) | S-FTM16 | 15.0 | 60.0 | >50 | 55.46 | concave |
| | | | 70.0 | >50 | 126.42 | convex |
| Central distances between the optical elements (at 70 K) | | | | | | |
| Lens 1 – Lens 2 | | 5.0 | GM2 – Lens 1 | | 160.0 | |
| Lens 2 – Focal plane | | 290.3 (including filter with n=1.45 and thickness = 5.0) | | | | |
| Input focal plane – Lens 1 | | 405.0 | | | | |

### 5.5.1 Tolerances

Tolerances were computed with the same assumptions used for the lens-triplet camera (Sect. 5.3.1). The manufacturing tolerances are summarized in Table 20. The surface figure is the rms wave-front irregularity of each lens surface, measured in reflection over a diameter of 50mm.

**Table 20** Manufacturing tolerances for the lenses of the guider camera

| Element | Manufacturing tolerances | | | | | |
|---|---|---|---|---|---|---|
| | Curvature | C-Thickness | Wedge | Internal decenter | Surface figure (including power) | Surface roughness |
| Lens 1 ($CaF_2$) | 0.5% | 0.2 mm | 60 arcsec | 0.02 mm | 60 nm rms | 5 nm rms |
| | 0.5% | | | | 60 nm rms | 5 nm rms |
| Lens 2 (S-FTM16) | 0.5% | 0.2 mm | 60 arcsec | 0.02 mm | 60 nm rms | 5 nm rms |
| | 0.5% | | | | 60 nm rms | 5 nm rms |

The camera assembly includes both lenses. The tolerances reported here refer to the positioning of the whole assembly relative to the rest of the optics (slit, GM mirrors and guider detector).

Table 21 Positioning tolerances of the whole guider camera.

| Element | Lateral shift | Shift along optical axis | Tilt relative to optical axis |
|---|---|---|---|
| Camera assembly | 1 mm | 0.3 mm | 10 arc-min |

Table 22 lists the stability tolerances of the lenses for the two cameras. These tolerances also represent the accuracy of a mechanical holder where the lenses are blindly positioned, i.e. a situation where the alignment does not foresee the possibility of separately adjusting the shifts, tilts and central distances of the lenses.

Table 22 Positioning tolerances of the lenses within the camera

| Element | Stability and alignment tolerances | | |
|---|---|---|---|
| | Lateral shift | Shift along optical axis | Tilt |
| Lens 1 ($CaF_2$) | 0.03 mm | 0.1 mm | 3 arc-min |
| Lens 2 (S-FTM16) | 0.03 mm | 0.1 mm | 3 arc-min |

### 5.5.2 Anti-reflection coating

All the surfaces must be coated to minimize reflection. The specification for the anti-reflection coating is R<1% (goal R<0.5%) per surface over the 1000nm - 2500nm wavelengths range. Worse performances (i.e. higher reflection) may be accepted over the following ranges of wavelength

 1350nm – 1470nm
 1830nm – 1950nm